\documentclass[%
 reprint,
superscriptaddress,
 amsmath,amssymb,
 aps,
floatfix,
]{revtex4-1}
\usepackage{graphicx}
\usepackage{amsmath}
\usepackage{subfig}
\usepackage{comment}
\usepackage{amsmath}
\usepackage{amssymb}
\usepackage[below]{placeins}
\usepackage{fancyhdr}
\usepackage{hyperref}
\usepackage{cleveref}

\begin{document}


\title[\texttt{achemso} demonstration]
{Transport properties of ultrathin YBa$_2$Cu$_3$O$_{7-\delta}$ nanowires: a route to single photon detection}

\author{Riccardo Arpaia}
\affiliation{Quantum Device Physics Laboratory, Department of Microtechnology and Nanoscience, Chalmers University of Technology, 41296 G\"{o}teborg, Sweden}
\author{Dmitri Golubev}
\affiliation{Low Temperature Laboratory (OVLL), Aalto University School of Science, P.O. Box 13500, FI-00076 Aalto, Finland}
\author{Reza Baghdadi}
\affiliation{Quantum Device Physics Laboratory, Department of Microtechnology and Nanoscience, Chalmers University of Technology, 41296 G\"{o}teborg, Sweden}
\author{Regina Ciancio}
\affiliation{CNR-IOM, TASC Laboratory, Area Science Park, Basovizza S.S. 14 km 163.5, I-34149 Trieste, Italy}
\author{Goran Dra\v{z}i\'{c}}
\affiliation{Laboratory for Materials Chemistry, National Institute of Chemistry, Hajdrihova 19, SI-1001 Ljubljana, Slovenia}
\author{Pasquale Orgiani}
\affiliation{CNR-SPIN, University of Salerno, I-84084 Fisciano (SA), Italy}
\author{Domenico Montemurro}
\affiliation{Quantum Device Physics Laboratory, Department of Microtechnology and Nanoscience, Chalmers University of Technology, 41296 G\"{o}teborg, Sweden}
\author{Thilo Bauch}
\affiliation{Quantum Device Physics Laboratory, Department of Microtechnology and Nanoscience, Chalmers University of Technology, 41296 G\"{o}teborg, Sweden}
\author{Floriana Lombardi}
\email{floriana.lombardi@chalmers.se}

\affiliation{Quantum Device Physics Laboratory, Department of Microtechnology and Nanoscience, Chalmers University of Technology, 41296 G\"{o}teborg, Sweden}
\date{\today}
\begin{abstract}
We report on the growth and characterization of ultrathin YBa$_2$Cu$_3$O$_{7-\delta}$ (YBCO) films on MgO (110) substrates, which exhibit superconducting properties at thicknesses down to 3 nm. YBCO nanowires, with thicknesses down to 10 nm and widths down to 65 nm, have been also successfully fabricated. The nanowires protected by a Au capping layer show superconducting properties close to the as-grown films, and critical current densities, which are only limited by vortex dynamics. The 10 nm thick YBCO nanowires without the Au capping present hysteretic current voltage characteristics, characterized by a voltage switch which drives the nanowires directly from the superconducting to the normal state. Such bistability is associated in NbN nanowires to the presence of localized normal domains within the superconductor. The presence of the voltage switch, in ultrathin nanostructures characterized by high sheet resistance values, though preserving high quality superconducting properties, make our nanowires very attractive devices to engineer single photon detectors.
\end{abstract}


\maketitle

\section{Introduction}
Single photon detectors, based on superconducting nanowires (SNSPDs) \cite{gol2001picosecond}, employ thin films with a thickness of few atomic layers, where the superconducting properties are well preserved after nanopatterning \cite{marsili2011single, natarajan2012superconducting, engel2015detection}.

At present, only conventional superconducors like NbN, NbTiN, Nb, or WSi,  where reliable nanofabrication routines are well established, have been employed to realize such devices. YBa$_2$Cu$_3$O$_{7-\delta}$ (YBCO), and more in general high critical temperature superconductors (HTS), are promising materials to be used as SNSPDs, since they meet all the crucial requirements for the detection, such as a short coherence length $\xi = \hbar v_F/\pi \Delta$ (where $v_F$ is the Fermi velocity and $\Delta$ is the superconducting gap) and, simultaneously, the highest possible transition temperature $T_C$ \cite{sobolewski1998ultrafast, parlato2013time, arpaia2015high, lyatti2016ultra}.  However, the serious degradation of the superconducting properties, which typically occurs in HTS thin films when the thickness approaches the nanometer scale, is a crucial issue for the realization of nanoscale devices.

The growth of ultrathin YBCO films has been object of intense research since the discovery of the cuprate superconductors. 
Initially, the focus was on few (ideally one) unit cell thick YBCO films, embedded between layers of PrBa$_2$Cu$_3$O$_{7-\delta}$ (PBCO), and the main systems under investigation - precursors of 2D superconductors \cite{balestrino2003superconductivity} - were YBCO/PBCO superlattices \cite{triscone1989ba, li1990interlayer} and PBCO/YBCO/PBCO trilayers structures \cite{terashima1991superconductivity, cieplak1994origin}.

The choice of using PBCO in combination with YBCO was initially made because of the peculiar PBCO properties, neither metallic nor superconducting, and with a crystallographic structure and lattice parameters very similar to YBCO. PBCO has been often used as a protective layer, since bare ultrathin YBCO films are characterized by degraded superconducting properties, as a consequence of the surface and chemical instability of the compound  \cite{terashima1991superconductivity}. However, the deposition at high temperature of a PBCO layer may also lead to some complications, like the interdiffusion at the interface between PBCO and YBCO, which brings to the substitution of Pr atoms in Y sites, with percentages as high as 30\% \cite{chan1993thickness}. It is known that a 45\% atomic interdiffusion into the first YBCO layer results in a $T_C$ of $\approx 30$ K for the resulting alloy \cite{balestrino2003superconductivity}. Thence, few layers YBCO thin films are characterized by a reduced $T_C$.

In this paper we show the growth and characterization of ultrathin YBCO films on MgO (110) substrates, which exhibit superconductivity at thicknesses down to 3 nm. High quality superconducting properties have been obtained by using a Au film as capping layer. Nanowires with widths down to 65 nm have been also realized on bare 10 nm thick films, and on films capped with Au. The YBCO nanowires without Au capping show a voltage switch in the current voltage characteristics (IVC), similarly to that of NbN nanowires, which brings the system directly from the superconducting to the normal state. This feature, occurring in  structures with high sheet resistance and high critical current density, makes our ultrathin ($\approx 10$ nm) nanowires ideal candidates for ultra fast single photon detection.

\section{HTS as materials for SNSPDs} \label{sec: new}

YBCO, and HTS in general, are of great interest in the quest for new materials for SNSPDs. 
The high critical temperature,  $T_C^{YBCO} \approx 90$ K, could broaden the ultralow temperature range of operation (1.7 - 5 K) characterizing traditional SNSPDs made of low critical temperature superconductors, such as NbN, NbTiN and Nb.
The coherence length, $\xi_0^{YBCO} \approx 2$ nm, comparable to that of traditional SNSPD materials, $\xi_0^{NbN} \approx 5$ nm \cite{semenov2009optical}, offers the possibility to squeeze the dimensions of the detecting elements down to several nanometers without entering in a regime dominated by phase fluctuations, where dark counts would play a big role \cite{kitaygorsky2007dark, bartolf2010current}. 
HTS SNSPDs can also be interesting for a more fundamental physics point of view, offering new scenarios for intrinsic mechanisms involved in the formation of the photoresponse signal. In LTS SNSPDs, models based on the vortex-assisted photon counting are nowadays the most commonly accepted to describe the detection process (although none of the models proposed so far has been able to describe all the experimental data) \cite{renema2014experimental, engel2015detection}. Superconductivity in HTS materials and nanostructures is however quite different from conventional superconductors: the role of vortices could for example be different, as a consequence of the peculiar vortex dynamics characterizing this class of materials \cite{blatter1994vortices}.

Several works have been devoted to the realization of YBCO nanowires for their application in single photon detectors \cite{curtz2010patterning, lyatti2016high, amari2016ion}: despite all this technological effort, a HTS SNSPD has still not been realized. The reason for this could lay on the different material properties of YBCO with respect to LTS materials, commonly used in SNSPDs.

\begin{figure}[hbpt!]
\includegraphics[width=0.47\textwidth]{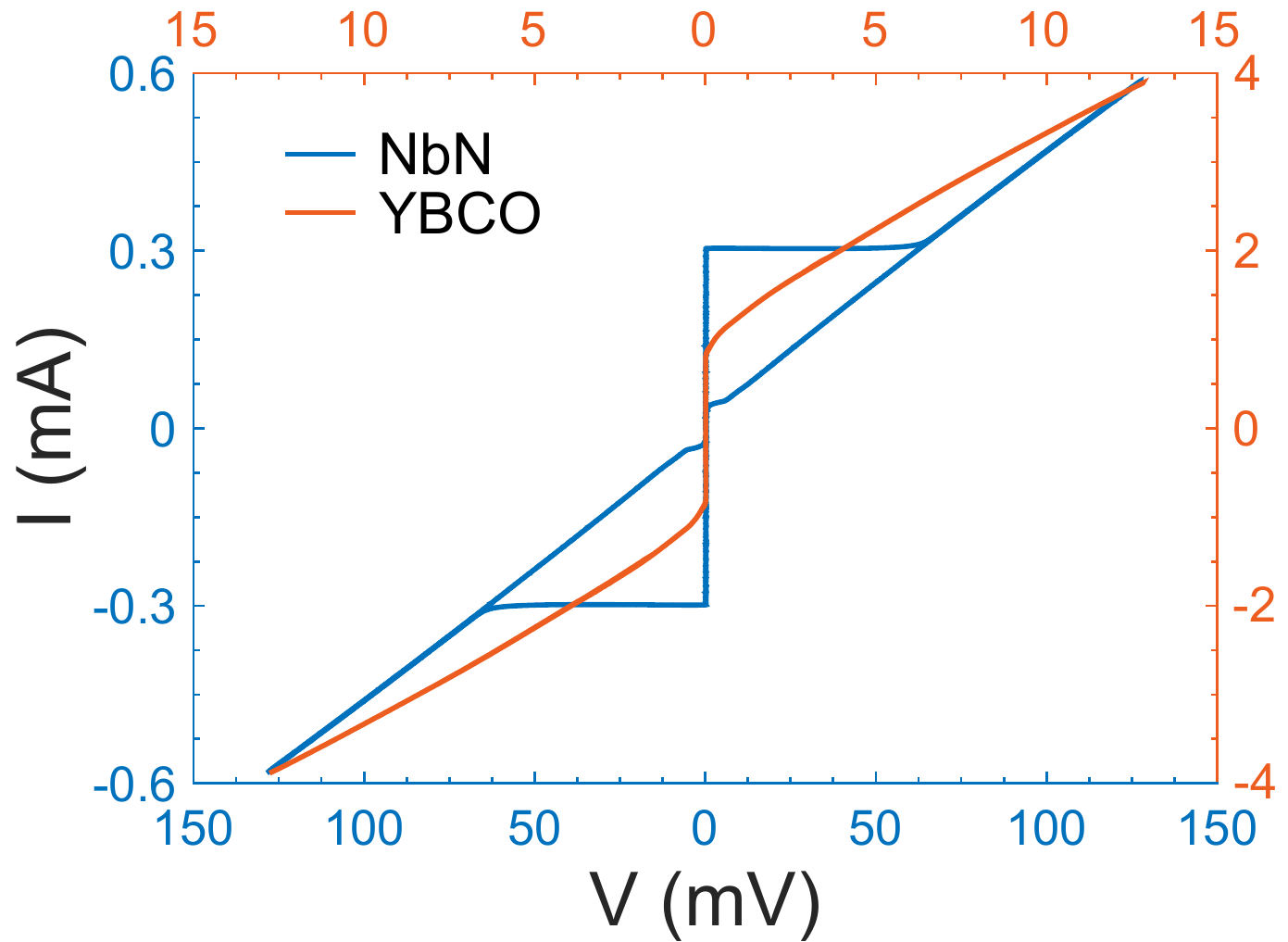}\\
\caption{\footnotesize Current voltage characteristics at $T = 4.2$ K of a 50 nm thick, 75 nm wide and 120 nm long NbN nanowire ({\itshape blue}) and of a bare YBCO nanowire ({\itshape orange}) having the same dimensions. Both the nanowires have been realized following the nanopatterning procedure described in Refs. \cite{arpaia2013improved, nawaz2013approaching}. In NbN, a large voltage switch, occurring just above the critical current, brings the whole nanowire into a state characterized by a differential resistance corresponding to the normal resistance $R_N$. In YBCO, a flux-flow behavior, characterized by differential resistances $\ll R_N$, dominates at currents higher than the critical one.} \label{Fig: Fig0}
\end{figure}

LTS nanowires are commonly characterized by high critical current densities, hysteretic IVCs and high sheet resistances. High critical current densities are required to improve the signal to noise ratios; the amplitude of the output signals is proportional to the superconductive critical current, which is limited by the low cross-sectional area of the wire. Hysteretic IVCs ensure the switching mechanism of a SNSPD as a consequence of the photon event. The presence of the hysteresis is very common in NbN nanostructures already at thicknesses, $t =50$ nm, much larger than those requested for their employment in SNSPDs (see Fig. \ref{Fig: Fig0}).
The hysteresis is a distinctive feature of the formation and development of a self-stabilizing hotspot in the superconducting wire \cite{il2005critical, stockhausen2012adjustment}, and it is a consequence of the low value of thermal conductivity in this material ($\kappa \approx 10^{-2} - 10^{-1}$ Wm$^{-1}$K$^{-1}$ in the normal state). Finally, high sheet resistance values make the nanowire resistance much higher that the external load resistance $R_L$ (typically 50 $\Omega$) in parallel with the device, so favoring the redistribution of all the bias current into the readout line, where the signal appears. In NbN, this is commonly achieved using microcrystalline, granular, ultrathin films: the normal resistance $R_N$ increases without reducing the homogeneity of the films, and nanodefects, acting as pinning centers, may even enhance the critical current density of the devices \cite{yurchenko2013thermo, lin2013characterization}.

For an ideal YBCO-based SNSPD, the critical current density $J_C$ should be as close as possible to the depairing/vortex entry limit, which equals to $J_v \approx 8 \cdot 10^7$ A/cm$^2$ at $T = 4.2$ K  \cite{bulaevskii2011vortex, arpaia2014resistive} (see section \ref{sec: Auwires} for more details). Until now, this condition     has been fulfilled only in nanowires having thickness of several tens of nanometers \cite{nawaz2013microwave, arpaia2015high}.
A first limitation, for the employment of YBCO nanowires with such thicknesses as single photon detectors, derives from the shape of the current voltage characteristic. Indeed the IVCs of YBCO nanowires with thicknesses of several tens of nanometers  are flux-flow like (see Fig. \ref{Fig: Fig0}) \cite{papari2014dynamics, amari2016ion}, presenting at most voltage switches in the resistive state (i.e. at currents well above the critical one), as a consequence of phase slips or of entry of Abrikosov vortices \cite{assink1993critical, mikheenko2005phase}. The high value of the thermal conductivity in YBCO, which is two orders of magnitude higher than in NbN \cite{sutherland2003thermal}, prevents self-heating effects that could bring the whole device from the superconducting directly to the normal state. Indeed the average differential resistance ${\delta}R$ in the measured finite voltage range of the IVC is much lower than the resistance $R_N$ of the nanowire, one measures at the onset of the superconducting transition. Therefore a photon event, independently of the detection mechanism which takes place, cannot drive the whole nanowire in the normal state, since ${\delta}R \ll R_L$, preventing a proper redistribution of the bias current from the wire into the load. Another drawback for the employment of YBCO in SNSPD comes from the impossibility to increase the sheet resistance of  the nanowires by increasing the granularity of the films. This is because grain boundaries significantly reduce the critical current instead of increasing it (as it happens for NbN): granular sub-100-nm-wide YBCO nanowires present $J_C$ not higher than $2 \cdot 10^6$ A/cm$^2$ \cite{levi2013periodic}.

This complex scenario has been confirmed by our previous works \cite{arpaia2014highly, arpaia2015high}, where we have shown that 50 nm thick YBCO nanowires, despite critical current densities $J_C$ close to the deparing limit, are characterized by a typical flux-flow like behavior in the full temperature range. They have shown a pure bolometric response under irradiation.

To overcome these limitations, one needs to employ ultrathin YBCO films. Indeed in films with thickness $t \le 15$ nm the YBCO resistivity increases  \cite{gao1995formation, tang2000thickness, poppe1992low}, therefore the thermal conductivity, via the Wiedemann-Franz law, decreases. Nanowires patterned on these films can be characterized by a physics close to that describing NbN nanowires, giving rise to hysteretic IVCs. The issue is thefore to reduce the thickness of YBCO nanowires, keeping the superconducting properties as close as possible to the bulk material. Indeed, as a consequence of the chemical instability of this material, together with its extreme sensitivity to defects and disorder, a significant $J_C$ drop may occurr at thicknesses comparable with $\xi$, reducing the Joule heating in the nanowires, therefore the probability to get hysteretic IVCs. Previous attempts, to realize ultrathin YBCO nanowires, have lead to critical currents, for the sub-100-nm-wide bridges, more than one order of magnitude lower than $J_v$: as a consequence, despite the small cross sections, the IVCs are still flux-flow like \cite{curtz2010patterning} or they show voltage switches only in the resistive state \cite{lyatti2016high}.  

The structure of the paper is as follows. In Sections \ref{sec: filmstruc} and \ref{sec: filmtrans} we will present the structural and transport characterization of ultrathin YBCO films. We will show that - despite the lack of PBCO encapsulation, commonly used to protect thin YBCO  - the film properties are intrinsic of the material, confined at thicknesses of few nanometers. In Section \ref{sec: Auwires} we will study Au capped YBCO   nanowires, with thickness down to 10 nm, to show that our nanopatterning procedure, already successfully used on 50 nm thick nanowires \cite{nawaz2013microwave}, can preserve the superconducting properties also in ultrathin structures. These nanowires are used as reference structures for the maximal $J_C$ value one can achieve at this thickness. Finally in Section \ref{sec: Cwires} we will focus on bare 10 nm thick YBCO nanowires, to show their potentialities as SNSPDs.

\section{Ultrathin films: growth and structural characterization} \label{sec: filmstruc}

We have deposited YBCO films, with thicknesses from 50 nm down to 3 nm, on MgO (110) substrates by Pulsed Laser Deposition (heater temperature 760 $^\circ$C, oxygen pressure 0.7 mbar, energy density 2 J/cm$^2$). After the deposition, the films have been slowly cooled down (cooling rate 5 $^\circ$C/min) at oxygen pressure of 900 mbar, to promote full oxidation of the YBCO chains. With this procedure, we have previously shown that 50 nm thick YBCO films are slightly overdoped \cite{baghdadi2015toward}. For each thickness, we have deposited two YBCO films: one with a 50 nm thick Au capping layer, in situ sputtered after the YBCO deposition, and another without protecting layer.

A morphological and structural characterization has been carried out by using Atomic Force Microscopy (AFM) and Scanning Electron Microscopy (SEM) to establish the quality of the YBCO films. The films present smooth surfaces, characterized by the typical $c$-axis domains with 3D spirals, and an average roughness which is of the order of one atomic cell (see Supplemental Material, sec. I).

\begin{figure}[hbpt!]
\includegraphics[width=0.37\textwidth]{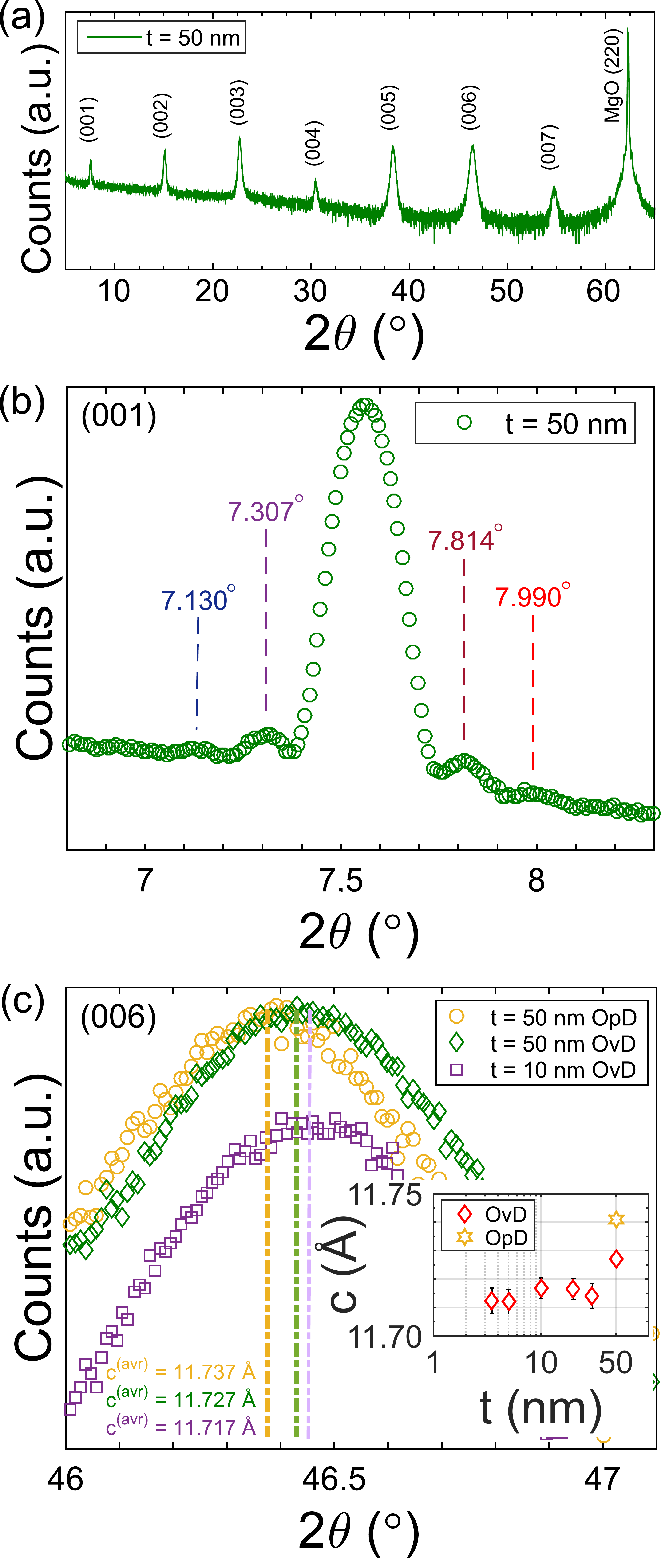}\\
\caption{ {\footnotesize a) $2\theta-\omega$ X-ray diffraction scan of a 50 nm thick slightly overdoped (OvD) YBCO film. b) $2\theta-\omega$ XRD scan of the (001) Bragg reflection. c) $2\theta-\omega$ XRD scans of the (006) reflections of a 10 nm (violet circles) and 50 nm (green circles) thick slightly overdoped YBCO films, respectively. A 50 nm thick nearly optimal doped YBCO film, obtained by using a reduced annealing oxygen pressure \cite{baghdadi2015toward}, has been also shown for comparison (OpD, orange circles). The error bars are within the experimental points in all the cases. The maximum of each peak is highlighted by the dashed line. The $c$-axis average values, $c^{(avr)}$, extracted considering all the (00$n$) reflections measured for the three films, are also reported. In the inset, $c^{(avr)}$ is plotted as a function of the film thickness.}
} \label{Fig: Fig1b}
\end{figure}

The structural properties have been determined by X-ray diffraction (XRD) analysis with
a 4-circle Panalytical X'pert diffractometer with CuK$_\alpha$ radiation, using a hybrid Ge(220) monochromator and a
PIXcel 3D detector matrix. Symmetric $2\theta-\omega$ scans show that the ultrathin films are highly crystalline and $c$-axis oriented  (see Fig. \ref{Fig: Fig1b}a). The crystallographic quality of the films is confirmed by the presence of the interference fringes which are visible on both sides of the lowest diffraction order (0 0 $n$) YBCO Bragg-peaks (see Fig. \ref{Fig: Fig1b}b), in films with thicknesses down to 20 nm. The thickness $t$ has been calculated, with an error of less than 1 atomic layer ($\approx 1$ nm), by considering couples of adjacent maxima and  using the relation $t = \lambda_{Cu} \cdot [2\cdot(\sin \omega_n - \sin \omega_{n-1})]^{-1}$ (where $\omega_{n}=2\theta_n/2$ are the angles corresponding to the maxima of the interference fringes and $\lambda_{Cu} = 1.540598$ \AA \, is the wavelength of the incident X-ray wave).

From the angular position of the  (00$n$) peaks we can estimate the $c$-axis length. 
On the basis of the bulk lattice parameters of YBCO and MgO (110), 
a large in plane lattice mismatch $\delta^m$ (defined as $| a_{\mathrm{sub}}-a_{\mathrm{film}}|/a_{\mathrm{sub}}$) is expected between the films and the substrate   ($\delta^m$ is $\approx 9$ \% and $\approx 35$ \% along the [0 0 1] and [1 -1 0] MgO directions respectively) \cite{zheng1992early}. In ultrathin YBCO films, especially when grown on substrates with a large mismatch, a $c$-axis expansion has been usually reported \cite{savvides1994growth}, and attributed to a large number of misfit dislocations caused by the extreme strain induced by the substrate on the whole film thickness. However in our films, in spite of the large mismatch between the films and the substrates, no $c$-axis expansion is measured.  In more detail, our films show a c-axis length of $\approx 11.715$ \AA, which is sizebly smaller than those we have measured in optimally doped films grown on MgO (110) substrates, corresponding to $\approx 11.74$ \AA  \, (see Fig. \ref{Fig: Fig1b}c, and inset therein). Interestingly, films below 30 nm thickness have a slightly shorter $c$-axis (i.e. $11.705$ \AA) than the reference slightly overdoped 50 nm thick films (see Fig. \ref{Fig: Fig1b}c, and inset therein). Such a small difference, compatible with a higher oxygen content, might be related to the efficiency of the post-annealing in incorporating oxygen in films of only few unit cells.

\begin{figure}[hbpt!]
\includegraphics[width=0.45\textwidth]{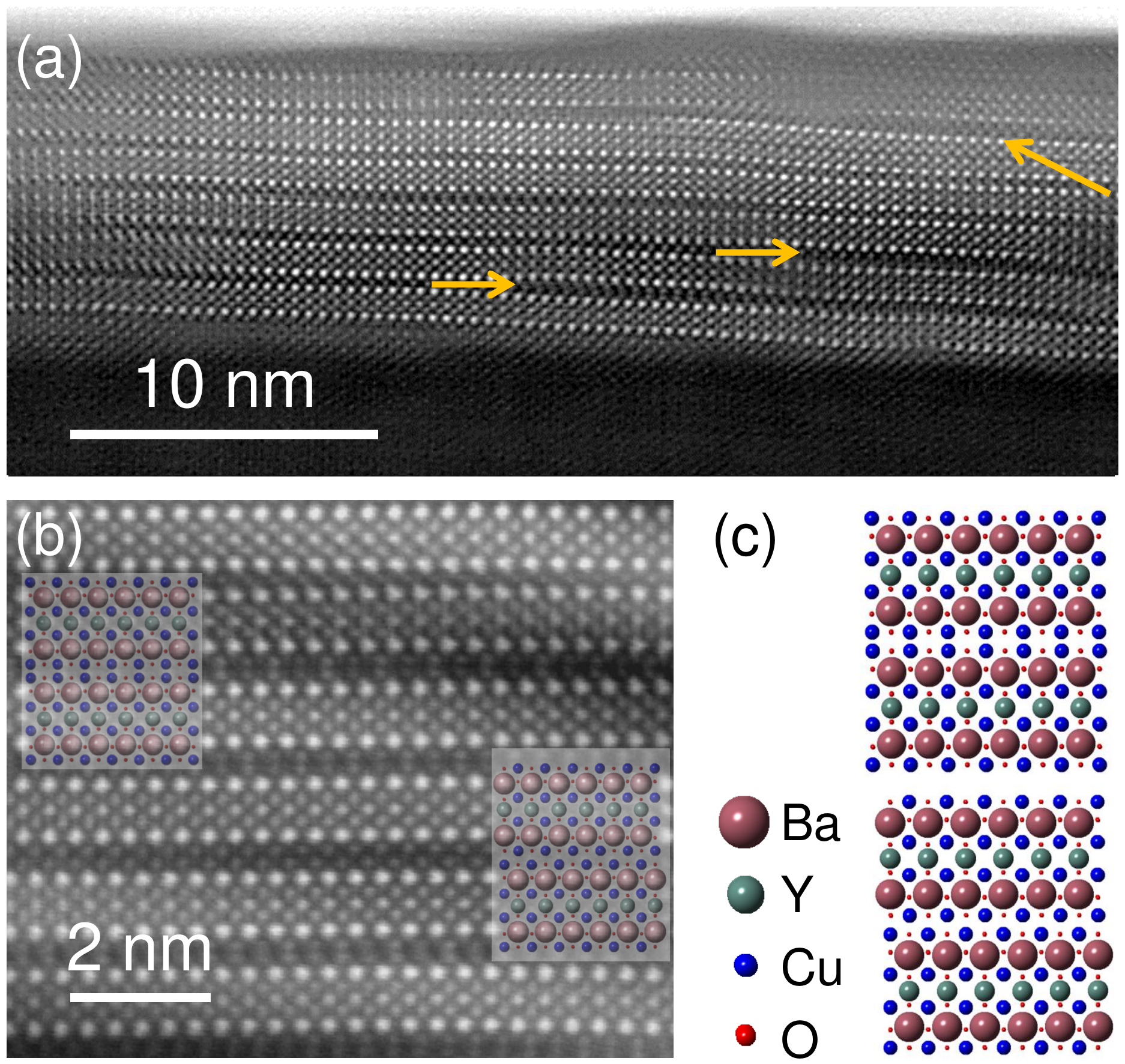}\\
\caption{a) Atomic resolution HAADF-STEM image of a representative cross-sectional region of a 10 nm thick film of YBCO. The presence of Y248 intergrowths is highlighted by arrows. b) Close up view showing the [010] and [100] zone axes of the Y248 crystal structure. The two orientations of the Y248 cell are shown in (c) and superimposed to the relative areas in (b).
} \label{Fig: Fig1a}
\end{figure}

Cross sectional High Resolution Transmission Electron Microscopy (HRTEM) and High Angle Annular Dark Field Scanning
TEM (HAADF-STEM) investigations performed on 10 nm thick films unveiled a complex nanostructure of the films characterized by a high density of Y$_2$Ba$_4$Cu$_8$O$_{16}$ (Y248) intergrowths intercalating within the YBCO matrix. Figure \ref{Fig: Fig1a}a shows a high resolution Z-contrast image of the YBCO film 
where the presence of Y248 planar faults is observed (indicated by arrows). In particular, as highlighted in the enlarged image (see Fig. \ref{Fig: Fig1a}b), the Y248 intergrowths are sometimes accompanied by a rigid shift of half the unit cell along the $b$-axis of YBCO; hence, one can easily identify the [010] and [100] zone axes of the Y248 structure as a consequence of YBCO twinning.  As a result of the presence of a high concentration of intergrowths, a waving of the (00$\ell$) planes is observed with a consequent break of the vertical coherence which extends over the whole film thickness. The same nanostructural assessment is observed in case of the 50 nm thick films (not shown here). The presence of Y248 planar faults in YBCO thin films has been previously reported \cite{guzman2013strain}.  

The Y248 intergrowths introduce additional CuO chains into the structure, making our thin films Cu-rich and overdoped at all thicknesses. This latter characteristic has been additionally confirmed by a transport characterization of the films, as it will be discussed in the next section.

\section{Ultrathin films: transport characterization} \label{sec: filmtrans}

\begin{figure*}[!bhpt]
\includegraphics[width=0.85\textwidth]{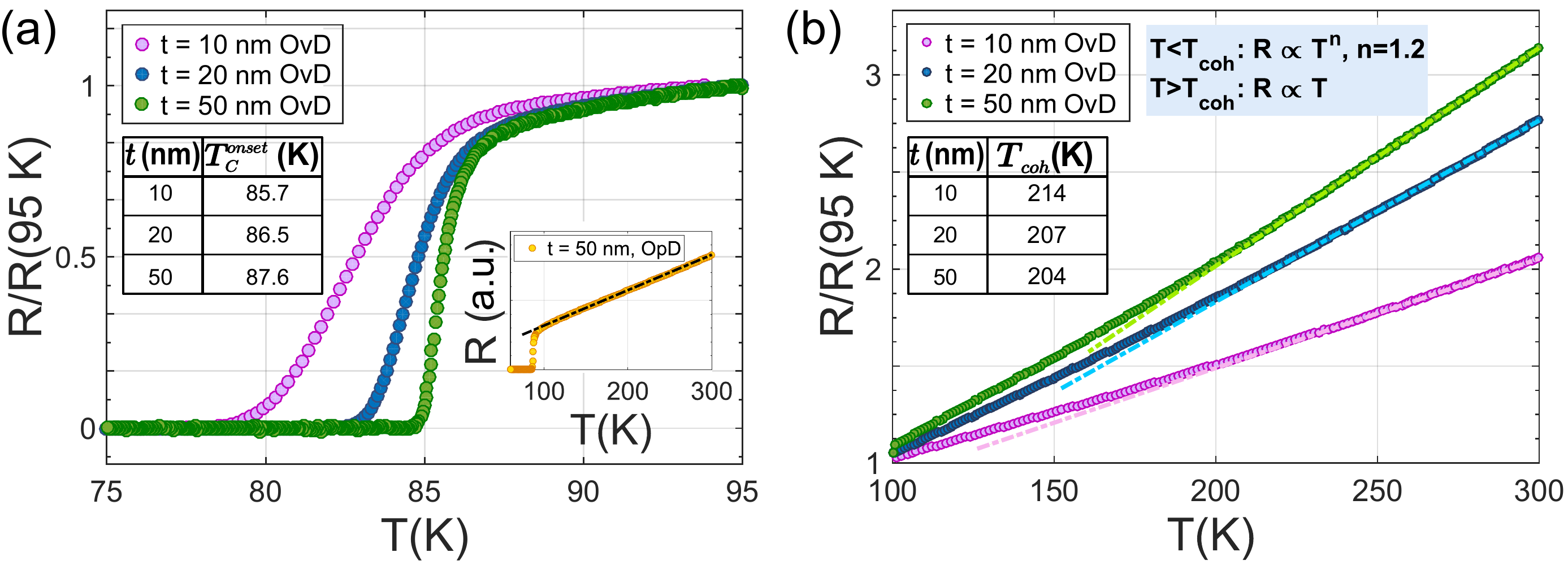}\\
\caption{ a) $R(T)$ curves, normalized to the value of the resistance at 95 K, of bare YBCO films with thicknesses $t = 10, 20$ and 50 nm. The $T_C^{onset}$ of the three films, listed in the inset table, are calculated as the temperature where the resistance $R = 0.9 \cdot R_N$ (with $R_N$ being the normal resistance). The inset shows the $R(T)$ measurement of a 50 nm thick optimal doped bare YBCO film ($T_C^{onset} = 89.1$ K), which is linear down to $\approx 105$ K. b) Normalized resistance of the same three films as panel a), in the temperature range between 100 and 300 K. The dashed lines represent the linear fit to the data at temperatures close to 300 K. The linear dependence is in good agreement with the data down to $T_{coh}$. Below $T_{coh}$, the $R(T)$ data can be fitted assuming a power law dependence $R = a + b \cdot T^n$, where $a$ and $b$ are free parameters, and $n = 1.2$ for the three films. The resulting $T_{coh}$ of the three films are listed in the inset table. The $R_{\square}$ values measured at 100 K in these three films are 130, 33 and 12 $\Omega$, respectively for 
$t =$ 10, 20 and 50 nm.
} \label{Fig: Fig2}
\end{figure*}

Transport properties of the ultrathin YBCO films have been studied by resistance vs temperature $R(T)$ measurements.  For ultrathin YBCO films, when grown on substrates with a large mismatch (without a PBCO seed layer), a strong reduction of  the zero resistance temperature, $T_C^0$, has been reported \cite{probst2012magnetoresistivity}. Here we show instead that the $T_C^0$ of our films is higher than the liquid nitrogen temperature down to thicknesses of 10 nm. From the onset of the superconducting transition and the normal state properties of the ultrathin films we are able to extract information about the oxygen doping of nanometer thick YBCO films.

\subsection{Determination of the oxygen doping}
In Figures \ref{Fig: Fig2}a and \ref{Fig: Fig2}b the $R(T)$ measurements of the bare 10 and 20 nm thick films are compared with those of the 50 nm thick film (also without Au capping). 

The onset temperature of the superconducting transition, $T_C^{onset}$, slightly changes with the film thickness (see inset table in   Fig. \ref{Fig: Fig2}a), while $T_C^0$ substantially decreases by reducing the thickness (due to the significant broadening of the superconducting transition (see Fig. \ref{Fig: Fig2}a).

In principle a lower $T_C^{onset}$ could be associated both to an increase or to a reduction of the oxygen doping within the film. In our case, we can attribute the $T_C^{onset}$ reduction to a better oxygenation of the ultrathin films. This is clear from the normal state $R(T)$ dependence of these films  (see Fig. \ref{Fig: Fig2}b) and from the comparison with the $R(T)$ of an  optimally doped 50 nm thick film (see inset of Fig. \ref{Fig: Fig2}a).

The optimally doped films are characterized by a linear behavior of the resistance as a function of the temperature in the full temperature range, down to $T_C^{onset}$, as expected at this oxygen doping \cite{hussey2008phenomenology} (see inset of Fig. \ref{Fig: Fig2}a). The $R(T)$ measurements of the other three films are instead linear only at high temperature, while they exhibit a power law dependence, with the same exponent $n = 1.2$, below a temperature $T \approx 200$ K. This behavior is typical of fully oxygenated YBCO films, lying in the overdoped side of the phase diagram. The overdoped region of HTS materials is characterized by the so-called coherence temperature $T_{coh}$ (a crossover temperature from a coherent and an incoherent metal state \cite{kaminski2003crossover}), which increases as the oxygen doping increases: below $T_{coh}$ the dependence of the resistance with the temperature is not linear any longer and it becomes a power law like, $R(T) = a + b \cdot T^n$. The exponent $n$ varies from 1 at the optimal doping to 2 at the oxygen doping corresponding to the disappearance of the superconducting phase \cite{hussey2008phenomenology, manako1992transport, mackenzie1996normal}. As shown in the inset table of Figure \ref{Fig: Fig2}b, the $T_{coh}$ slightly increases by 10 K, when reducing the film thickness to 10 nm, signifying an oxygen content at least comparable with 50 nm thick films, if not slightly higher than that.

The $R(T)$ of the films has been measured with the Van der Pauw method \cite{van1958method}, to extract the sheet resistance $R_{\square}$ and the resistivity $\rho = R_{\square}{\cdot}t$, and to determine their dependence on the film thickness. The resistivity is thickness independent in films down to 20 nm, with a value $\rho \approx 60\,\mu\Omega\cdot$cm at $T = 100$ K, which is comparable with that obtained from YBCO single crystals \cite{liang1992growth}. In 10 nm thick films 
$\rho$(100 K) is instead $\approx 130\,\mu\Omega\cdot$cm: such increase is expected, since in literature the $R_{\square}$ is reported to scale faster than $1/t$ in sub-15 nm thick films  \cite{gao1995formation, tang2000thickness, poppe1992low}, where a significant drop of $T_C^0$ occurs (see next subsection).

\subsection{Broadening of the superconducting transition}

\begin{figure}[!h]
\includegraphics[width=0.48\textwidth]{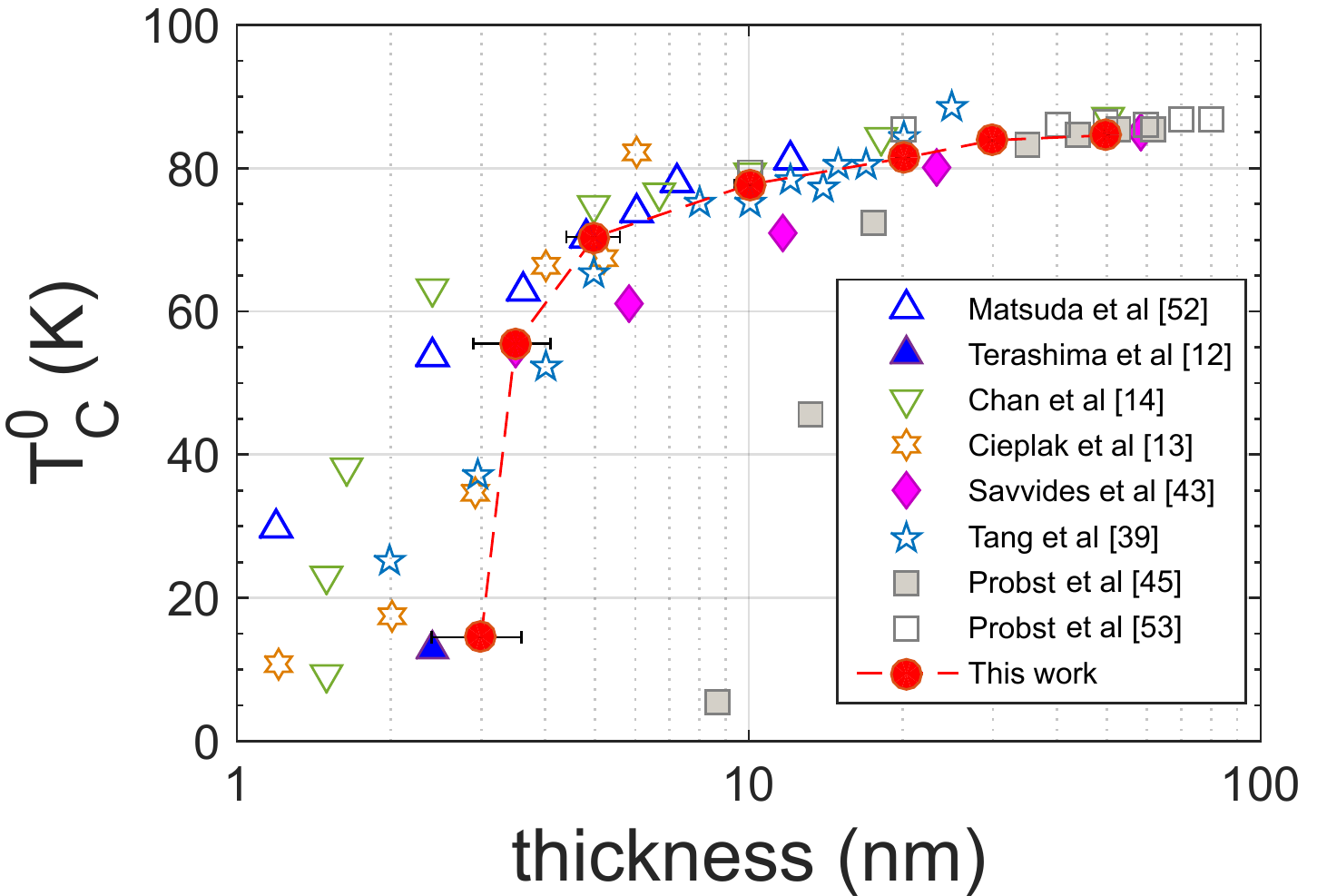}\\
\caption{Zero resistance critical temperature $T_C^0$ of our capped and bare YBCO films versus their thickness (red circles). Our results have been compared to those previously achieved in literature. In particular, open symbols show results achieved on YBCO films sandwiched between two PBCO layers by Matsuda et al \cite{terashima1991superconductivity, matsuda1993thickness}, Chan et al \cite{chan1993thickness}, Cieplak et al \cite{cieplak1994origin}, Tang et al \cite{tang2000thickness}, and Probst et al \cite{probst2012nonthermal}. Filled symbols show instead results achieved on YBCO films, not embedded between two PBCO layers, by Terashima et al \cite{terashima1991superconductivity}, Savvides et al \cite{savvides1994growth}, and Probst et al \cite{probst2012magnetoresistivity}.} \label{Fig: Fig3}
\end{figure}

In Figure \ref{Fig: Fig3} we have plotted (red circles) the $T_C^0$ extracted by our YBCO films - bare films above $t = 10$ nm and Au capped films below  $t = 10$ nm - as a function of the thickness.
As discussed earlier the $T_C^0$ slightly decreases down to 5 nm, which corresponds approximately to 4 unit cells. For lower thicknesses, it abruptly drops. Films with thickness of 3 nm, where only two complete atomic layers are present, are still superconducting with a $T_C^0 = 14.5$ K. At even lower thicknesses the films are not superconducting down to 4.2 K.  

We have compared our results with those published in literature starting from the beginning of the '90s up to now. 

As previously mentioned, most of the works have focused on YBCO films sandwiched between two PBCO layers \cite{terashima1991superconductivity, matsuda1993thickness, chan1993thickness, cieplak1994origin, tang2000thickness, probst2012nonthermal}. The $T_C^0$ values as a function of the YBCO thickness, collected from these papers, are shown as open symbols in Figure \ref{Fig: Fig3}. In all the previous works, $T_C^0$ gradually decreases down to 4 unit cell thick ($t \approx 5$ nm) YBCO films, while presenting a sudden drop at lower thicknesses. The same behavior can be also seen in superlattices, when the YBCO layers are interchanged either with PBCO layers \cite{varela1999intracell} or SrTiO$_3$ layers \cite{garcia2013disorder}. Below 5 nm thick films, the spread of $T_C^0$ values, which have been measured in different works, becomes wider. The origin of this spread can have several possible reasons, related to an error  in the determination of the thickness and/or to  small differences between different samples (in the interdiffusion rate between Pr and Y atoms at the interfaces or in the PBCO stoichiometry and/or the presence of crystalline defects) \cite{cieplak1993submonolayers}. 

Our films ({\itshape filled circles}  in Figure \ref{Fig: Fig3}), capped at most with a Au film, are characterized by $T_C^0$ values which are comparable, down to 5 nm, with the best results of PBCO/YBCO/PBCO multilayers ({\itshape open triangles}), while the $T_C^0$ of the very thin films are the best reported values for bare YBCO structures.


The wide broadening of the superconducting transition (defined as $T_C^{onset} - T_C^0$), significantly suppressing the $T_C^0$ of nanometer size thin films, has been commonly attributed to an intrinsic size effect of the material, and explained in terms of the Kosterlitz-Thouless (KT) transition related to vortex-antivortex pair dissociation in 2-dimensional systems \cite{beasley1979possibility, matsuda1993thickness, triscone1997superlattices}. Even though such interpretation has not achieved a general consensus \cite{cieplak1994origin, repaci1996absence}, other mechanisms implicating transfer of charge carriers from the YBCO to the PBCO layer as well as the presence of structural defects at the PBCO/YBCO interfaces \cite{cieplak1993submonolayers} can be ruled out in our case, since we are dealing with bare YBCO films (or capped with Au for thicknesses below 10 nm). 

\begin{figure*}[!htbp]
\includegraphics[width=0.96\textwidth]{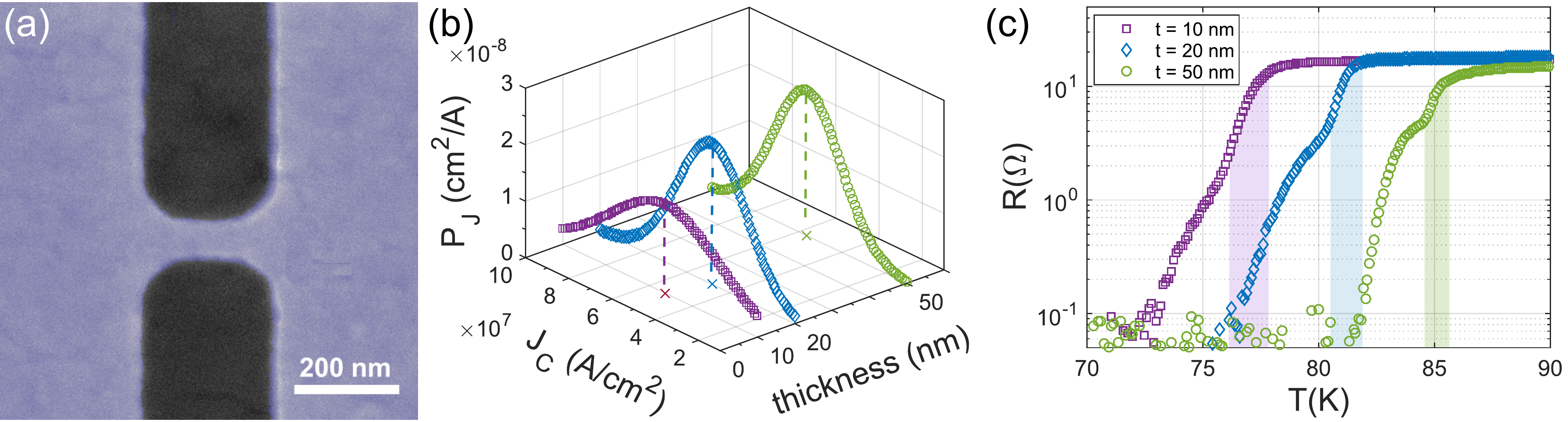}\\
\caption{ a) Scanning Electron Microscopy (SEM) picture of a Au capped 10 nm thick and 65 nm wide YBCO nanowire. b) Distribution $P_J$ of the critical current densities $J_C$ at T = 4.2 K of the population $\Pi$ of 65 nm wide Au capped nanowires, as a function of the wire thickness  ($\Pi_{10nm}=12$, $\Pi_{20nm}=59$ and $\Pi_{50nm}=18$). The $\bar{J}_C$ values of the three distributions, highlighted by the dashed lines, are almost identical ($\bar{J}_{C \, 10nm} = 5.3 \cdot 10^7$ A/cm$^2$, $\bar{J}_{C \, 20nm} = 4.8 \cdot 10^7$ A/cm$^2$ and $\bar{J}_{C \, 50nm} = 5.6 \cdot 10^7$ A/cm$^2$).  c) $R(T)$ of three 65 nm wide Au capped nanowires, at different thicknesses. The filled areas represent the temperature intervals between the $T_{C, w}^{onset}$ and  $T_{C, e}^{onset}$.
} \label{Fig: Fig4}
\end{figure*}

A second scenario, which could explain both the drop of $T_C^0$ and the raise of resistivity exhibited by the ultrathin films, points toward an increased disorder level in these systems, near the superconductor-insulator transition. To quantify the degree of disorder in our YBCO films as a function of the thickness, the product of the Fermi wave vector $k_F$ and electronic mean free path $\ell$ is commonly used \cite{ioffe1960non}. In YBCO, $k_F = 1.2{\cdot}10^9$ m$^{-1}$, while $\ell$ can be determined from the Drude model as $\ell = \hbar^2 v_F{(\hbar\omega_P)^{-2}}{(\epsilon_0\rho)^{-1}}$
where $v_F = 1.4{\cdot}10^5$ m$\cdot$s$^{-1}$ is the Fermi velocity, $\hbar\omega_P = 1.1$ eV is the plasma frequency \cite{gurvitch1987resistivity} and $\epsilon_0$ is the vacuum permittivity. In the 10 nm thick films, as a consequence of the increase of resistivity, the $k_F\ell$ product drops from $\approx 11.5$, value of  all the thicker films,  to $\approx 5.3$. In other superconductors, as NbN, similar $k_F\ell$ values have been associated to a moderate disorder, where a weakening of the pairing interactions is present but the role of phase fluctuation is still very limited \cite{chand2012phase}. The disorder becomes much stronger, and the phase fluctuations acquire a dominant role, in films with thickness $t < 10$ nm. Here, we cannot make a quantitative study, since the measured resistivities are dominated by the Au shunt. However, if we consider the same $\rho(t)$ dependence previously measured in other works \cite{gao1995formation}, the $k_F\ell$ products are expected to drop even further,  and the films to be on the edge of a superconductor-insulator transition ($k_F\ell \approx 1$). 

According to the latter scenario, the YBCO films with thickness $t \le 10$ nm are characterized by the same effective disorder, i.e. by the same $k_F\ell$ products, of NbN films used for SNSPDs \cite{kozorezov2015quasiparticle}.



\section{Ultrathin Au capped nanowires: pristine nanostructures} \label{sec: Auwires}

Nanowires with width down to 65 nm have been fabricated  (see Fig. \ref{Fig: Fig4}a) by using 10 and 20 nm thick slightly overdoped YBCO films both with and without Au capping, via a gentle Ar$^+$ ion milling, through an e-beam lithography defined hard carbon mask (for more details of the nanopatterning procedure, see Ref.  \citenum{arpaia2013improved, nawaz2013approaching}). 
The edges between the wires and the wider electrodes have been designed with a rounded shape, to minimize current crowding effects, which could be a source of reduction of the critical current density $J_C$ \cite{clem2011geometry, nawaz2013microwave}.

The nanodevices have been characterized via current-voltage characteristics (IVC) and resistance versus temperature $R(T)$ measurements. We have compared the results with those obtained on nanowires with the same geometry, obtained on 50 nm thick YBCO films capped with Au that we use as reference \cite{arpaia2013improved, nawaz2013approaching}. 

All the measured nanowires were superconducting. The IVC of the nanowires at $T = 4.2$ K exhibit a typical flux flow like behavior. From the IVCs we can determine the critical current $I_C$ of the nanowires, and the critical current density $J_C = I_C/(w \cdot t)$, where $w$ is the width of the nanowire.

To test the reproducibility of our nanostructures, we have measured, for each sample, a large number $\Pi$ of identical wires.  Figure \ref{Fig: Fig4}b shows the normal profile, determined by the distribution of the extracted $J_C$ values, as a function of the film thickness. For each distribution, we can extract
an average critical current density $\bar{J}_C$. The $J_C$ distributions become broader for thinner nanowires, possibly due to a most prominent role of defects at such reduced dimensions. However 
$\bar{J}_C$ is the same as for the 50 nm thick nanowires, and the highest $J_C$ within each distribution is $J_C^{max} \approx 8 \cdot 10^7$ A/cm$^2$. For nanowires which are characterized by dimensions $w, l \gg 4.44 \xi_0$ (where $\xi_0 \approx 2$ nm for YBCO is the superconducting coherence length at zero temperature), the maximum $J_C$ is given by the entry of Abrikosov vortices, driven by the Lorentz force, overcoming the bridge edge barrier. In this regime $J_v$ (critical current density due to vortex entry) can approximately be written as \cite{bulaevskii2011vortex, arpaia2014resistive}
\begin{equation}
J_{v} \approx 0.826 J_d \; ,
\end{equation}
where $J_d$ is the depairing critical current density, predicted by the Ginzburg-Landau theory, whose expression \cite{tinkham1996introduction},
\begin{equation}
J_{d} = \frac {\Phi_0}{ 3^{3/2}\pi\mu_0 \lambda_0^2\xi_0 } \label{Eq:J_d} \; ,
\end{equation}
equals to $1 - 3 \cdot 10^8$ A/cm$^2$ for YBCO nanowires \cite{arpaia2013improved, nawaz2013approaching, nawaz2013microwave}. In Eq. \ref{Eq:J_d}, $\Phi_0$ is the flux quantum, $\mu_0$ the vacuum permeability and $\lambda_0$ the London penetration depth at zero temperature. The maximum $J_C$ value of our nanostructures are therefore close to the theoretical limit, showing that  superconducting properties close to the as-grown films are preserved even at ultrathin thicknesses. 

The resistance as a function of the temperature close to the transition is shown in Figure \ref{Fig: Fig4}c for three nanowires, with thicknesses $t = 10, 20$ and 50 nm. For each device, the first transition in the $R(T)$, at higher temperature, is related to the wide electrodes connected to the nanowire: the temperature range of this transition corresponds to that of the unpatterned film (previously seen in Fig. \ref{Fig: Fig2}a), though the onset $T_{C, e}^{onset}$ is lower because of the shunt effect of the Au layer \cite{arpaia2013improved}. The second, broader, transition in the $R(T)$, at lower temperature, is instead associated to the nanowire: what is more remarkable of this transition is that the onset temperature $T_{C, w}^{onset}$ is typically only 1-2 K lower than the one of the wide electrodes, $T_{C, e}^{onset}$, independently of the thickness. 

The broadening of the superconducting transition of the thinnest nanowires is not much wider than that of the 50 nm thick nanowires. However, this broadening cannot be only ascribed to the dissipation induced by Abrikosov vortices crossing the nanowires, shown for the 50 nm thick nanodevices in other works \cite{arpaia2014resistive, arpaia2014yba, baghdadi2015fabricating}. Indeed, for ultrathin nanowires the picture is more complicated, since the broadening due to the thickness confinement, related to the various mechanisms mentioned in the previous section, including the KT transition, valid for 2D superconductors \cite{beasley1979possibility}, has to be taken into account. Thence, a quantitative analysis of the superconducting transition of ultrathin nanowires requires further investigations, which go beyond the scope of this paper.

\section{Ultrathin bare nanowires: a possible platform for YBCO single photon detectors} \label{sec: Cwires}

For most applications requiring ultrathin superconducting devices, a Au capping layer is not ideal. For single photon detection applications, for example, the high conductivity of Au leads to a high reflection of the light. At the same time, the Au capping layer works as an electrical shunt for the photoresponse and could also represent a thermal shunt for the perturbed area/hot spot inside the detector \cite{arpaia2015high}. 


The occurrence of hysteretic IVCs, associated to the formation of a self-stabilizing hotspot in a superconducting nanowire, is not very likely to happen in YBCO nanowires. As already mentioned in section \ref{sec: new}, this is mainly because of the high value of the thermal conductivity $\kappa$ in YBCO, in particular close to the optimal doped regime \cite{sutherland2003thermal} ($\kappa \approx$ 15 Wm$^{-1}$K$^{-1}$ in the normal state), preventing self-heating effects that could turn normal the central part of the nanowire.

Here we present the fabrication of bare YBCO nanowires, patterned on 10 nm and 20 nm thick films as those shown in secs. \ref{sec: filmstruc}-\ref{sec: filmtrans}. With respect to the nanofabrication procedure described in other works  \cite{arpaia2013improved, nawaz2013approaching}, a hard carbon mask - deposited by PLD at room temperature and removed at the end of the nanopatterning by oxygen plasma etching - was the only protecting capping layer for the YBCO film during the nanopatterning. The carbon layer protects YBCO from chemicals and from the direct impact of the Ar$^+$ ions during the etching. Because of its low thermal conductivity, it is not as effective as Au in preventing heating of YBCO during the baking of the resists, and during the ion milling. This may lead to possible deoxygenation, from the nanowire sidewalls during the nanopatterning, and the occurrence of some randomly distributed inhomogeneities.

Transport measurements have been done on 20 nm thick bare YBCO nanowires, characterized by resistivity values very close to those of 50 nm thick wires. From the IVCs at $T = 4.2$ K of the nanostructures, we have extracted $J_C$ up to  $5 \cdot 10^7$ A/cm$^2$, the average critical current density of the reference Au capped nanowires (see sec. \ref{sec: Auwires}). However, a voltage switch from the superconducting to the normal state has never been observed.


In the following we show that bare nanowires, with thickness $t = 10$ nm, are instead characterized, similarly to NbN nanowires, by hysteretic IVCs, with a voltage switch driving the structures abruptly from the superconducting to the normal state as soon as the critical current is exceeded. 

Current-voltage characteristics and resistance versus temperature measurements have been carried out on 38 nanowires, having widths in the range 65 - 120 nm, lengths in the range 100 - 500 nm, and patterned on several 10 nm thick films. 

\begin{figure*}[h!tbp]
\includegraphics[width=0.8\textwidth]{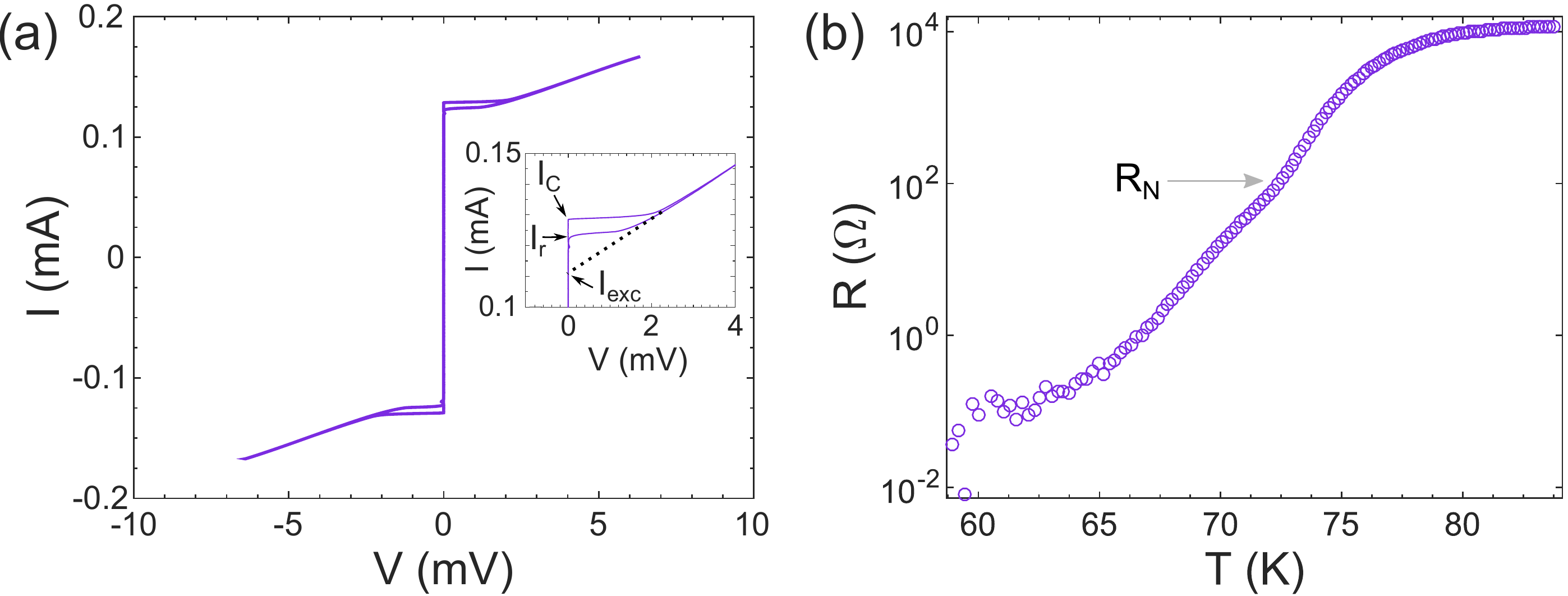}\\
\caption{a) The IVC measured at $T = 4.2$ K of a 75 nm wide and 100 nm long bare nanowire ($J_C = 1.8 \cdot 10^7$ A/cm$^2$) presents a 2 mV wide voltage switch above $I_C$, which drives the system to the normal state, corresponding to a differential resistance $dV/dI = 112$ $\Omega$ (see inset). The IVC present an excess current $I_{exc}$, possibly deriving from Andreev reflections at the interface between the superconductor and the normal region originating from a hot spot \cite{blonder1982transition}. b) $R(T)$ measurement of the same nanowire. The value of the normal resistance, corresponding to the $T_C^{onset}$ of the nanowire, is $R_N \approx 100$ $\Omega$, very close to the $dV/dI$ value measured in the IVC.
} \label{Fig: Fig5}
\end{figure*}

From the IVCs, measured at $T = 4.2$ K in current bias mode, we have inferred critical current densities as high as $J_C \approx 3.5 \cdot 10^7$ A/cm$^2$, quite close to the average critical current density of the Au capped nanowires (which are reference nanowires). As a consequence of the choice of carbon as a capping layer, we have obtained only a slightly lower $\bar{J}_C$,
but with  larger spread of $J_C$ values (see Supplemental Material, sec. II), and broader resistive transitions, compared with nanowires capped with Au (see Fig. \ref{Fig: Fig5}b). 

%


The IVCs of the nanowires with $J_C$ lower than  $\approx 5 \cdot 10^6$ A/cm$^2$, are flux flow like (see Supplemental Material, sec. II). The shape of the IVCs instead abruptly changes for nanowires with $J_C$ higher than this threshold. A large voltage switch, of the order of several mV, appears above $I_C$, which drives the devices to a state characterized by a constant differential resistance, $dV/dI \approx 1 - 4 \cdot 10^2 \, \Omega$, up to $V \approx 20$ mV, when applying a bias current $I_b > I_C$ (see Figs. \ref{Fig: Fig5}a). The difference between the switching ($I_C$) and the retrapping ($I_r$) current is in the order of several $\mu$A. The fact that we observe the voltage switch only in nanowires with the highest $J_C$, approaching the maximum current density a YBCO nanowire can carry, is an indication that the switching is not caused by the presence of local grain boundaries or defects.


The normal resistance $R_N$ (value of the resistance at the onset of the superconducting transition) of the nanowires is very close to the differential resistance $dV/dI$ measured in the IVC at 4.2 K in all the investigated samples (see Fig. \ref{Fig: Fig5}b), in agreement with what is observed in NbN nanowires (see Fig. \ref{Fig: Fig0}). This occurrence supports the hypothesis that the voltage switch above $I_C$ is a sign of self-heating in the nanowires: a normal/conducting domain, induced inside the ultrathin wire, drives the system directly from the superconducting to the normal state. This hypothesis is further supported by the presence of an excess current in the hysteretic IVC of the nanowires, again in accordance with results on NbN nanowires  (see Fig. \ref{Fig: Fig0}). The excess current, corresponding to the extrapolation to zero voltage of the normal branch at high bias, derives from Andreev reflections at the two interfaces between the superconductor and the normal region originating from a self-stabilizing hot spot \cite{blonder1982transition}. 

It is worth mentioning that the self-heating hotspot model is simplistic, and it is not meant to describe the detection mechanism in these nanowires: in type-II superconductors, like NbN and YBCO,  magnetic vortices play an important role (as already highlighted in sec. \ref{sec: Auwires}) \cite{bulaevskii2011vortex}; in addition to this, the physics of ultrathin films and nanowires is characterized by thermal and quantum phase slips \cite{zgirski2005size, astafiev2012coherent}. 
Nevertheless, NbN nanowires still work as SNSPDs whether their physics is dominated by phase slips or vortices \cite{delacour2012quantum, engel2015detection}. 
The understanding of the processes responsible for the formation and development of the normal/conducting domain within our nanowires upon irradiation goes beyond the purpose of the present paper.

Qualitatively, within the simplistic framework of the self-heating hotspot model \cite{skocpol1974self}, which has been even recently used to describe the IVC of NbN \cite{adam2010discontinuous, stockhausen2012adjustment, schmidt2017aln} and cuprate HTS \cite{golubev2014effect, Sophie} nanowires, the central part of the wire becomes normal due to Joule heating, producing a  power equal to $I\cdot V$. A cooling mechanism follows, via phonons and electron diffusion. The occurrence of the switch in the IVCs of 10 nm thick wires might be related to the increase of the film resistivity in YBCO below 15 nm \cite{gao1995formation, tang2000thickness, poppe1992low}.  As mentioned in sec. \ref{sec: filmtrans}, in our 10 nm thick nanowires we have measured at $T_C$ a resistivity $\rho$, which is more than a factor 2 higher than the $\rho$ of our 50 nm and 20 nm thick nanowires. Indeed an increase of resistivity would result in an increase of Joule heating, while at the same time the thermal conductivity $\kappa$ is decreased via the Wiedemann-Franz law, $\kappa (T) = \mathcal{L} T / \rho$, where $\mathcal{L} = \pi^2k_B^2/3e^2 \approx 2.45 \times 10^{-8}$ W$\Omega/$K$^2$ is the Lorentz number.  Both factors would support the stabilization of a hotspot in our ultrathin nanowires. It is worth mentioning that in NbN wires, the  value of the retrapping current $I_r$, that within the self-heating hotspot model depends on the $\rho$ of the material, on the thickness and on the thermal coupling between the wire and the substrate, has been experimentally found to decrease when decreasing the thickness \cite{schmidt2017aln}. This implies that the stabilization of a hotspot is favoured in ultrathin wires.

\begin{figure}[hbpt!]
\includegraphics[width=0.40\textwidth]{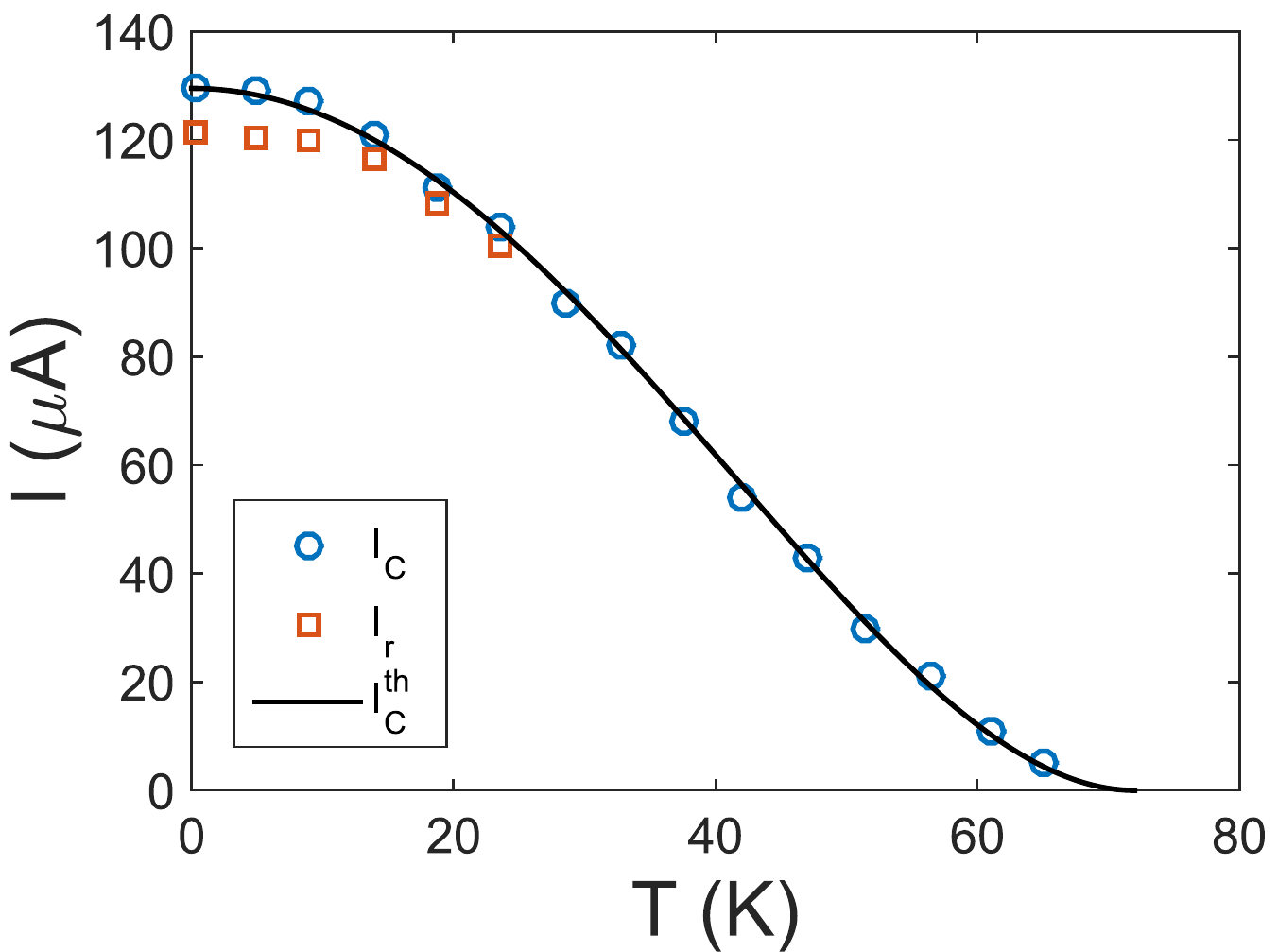}\\
\caption{Experimental critical current $I_C$ ({\itshape circles}) and retrapping current $I_r$ ({\itshape squares}) as a function of the temperature, for the nanowire shown in Figure \ref{Fig: Fig5}. The solid line is a guide for the eye.} \label{Fig: Fig6}
\end{figure}

Finally, we have measured the IVCs of the 10 nm thick bare YBCO nanowires as a function of the temperature, up to $T_C$:  the switching current $I_C$ decreases more rapidly than the retrapping current $I_r$ when increasing the temperature, leading to a gradual narrowing of the hysteresis (see Fig. \ref{Fig: Fig6}). 
In particular, the hysteresis is almost negligible if $T > 0.4\cdot T_C$. In NbN nanowires the hysteresis, which also decreases when increasing the temperature \cite{adam2009stabilization}, can instead persist up to temperatures close to $T_C$. In our case, the narrower $T/T_C$ range in which $I_C > I_r$ might be due to the smaller value of $\kappa$ of YBCO with respect to that of NbN.  This is also evident from the shape of IVCs at $T = 4.2$ K, presenting hysteresis, i.e. values of the retrapping voltage and difference between $I_C$ and $I_r$, much smaller than those of NbN nanowires (see Fig. \ref{Fig: Fig0}).

\section{Summary and conclusions}
We have shown the realization of YBCO films, which exhibit superconductivity down to 3 nm. The films show a  $c$-axis,  close to the reference 50 nm thick films, still corresponding to a slightly overdoped regime, and signifying a high crystallinity down to 2 -3 unit cells. Moreover, the $T_C^0$ of the thinnest films, obtained by using a Au capping, is comparable, down to 4 unit cells, with the best value obtained by sandwiching YBCO films with PBCO. Since the presence of the PBCO capping and buffering induces substitutions and/or chemical modifications, our results can be considered a first step toward the realization of few unit cell structures with properties representative of the YBCO bulk.

We have also fabricated YBCO nanowires with thicknesses down to 10 nm and widths down to 65 nm, both with and without the protection of a Au capping layer. The  ultrathin Au capped YBCO nanowires ($t = 10$ nm) behave similarly to the 50 nm thick nanowires, and are characterized by superconducting properties not affected by the nanopatterning procedure up to their $T_C^0$, a value   close to the $T_C^0$ of the unpatterned films. The IVCs are non-hysteretic. The ultrathin 10 nm thick bare YBCO nanowires are instead characterized by hysteretic current voltage characteristics, with a voltage switch, of the order of a few millivolts, from the superconducting to the normal state. In addition to this bi-stability of the IVC around $I_C$, in our bare nanowires coexist 1) very small cross sections, with sub-100 nm widths and thicknesses not far from the superconducting coherence length; 2) high critical current densities, approaching the critical depairing value, limited by vortex entry; 3) enhanced sheet resistances, expected in sub-15-nm-thick films and nanostructures, which make it possible to engineer sub-$\mu$m long wires having $k\Omega$ resistance values, which are much larger than the 50 $\Omega$ resistance used in parallel with the device in high speed SNSPD readout circuits.  The coexistence of these four features makes our ultrathin  YBCO nanowires very attractive for SNSPDs.

\vspace{1cm}

\begin{acknowledgements}
This work has been partially supported by the Swedish Research Council (VR) and the Knut and Alice Wallenberg Foundation.
\end{acknowledgements}
\bibliography{biblio}

\newpage
\widetext

\begin{center}
\textbf{\Large Supplemental Material}
\end{center}
\setcounter{equation}{0}
\setcounter{figure}{0}
\setcounter{table}{0}
\setcounter{section}{0}
\makeatletter
\renewcommand{\theequation}{S\arabic{equation}}
\renewcommand{\thefigure}{S\arabic{figure}}

\section{Ultrathin film morphology}

\begin{figure}[!ht]
\includegraphics[width=15cm]{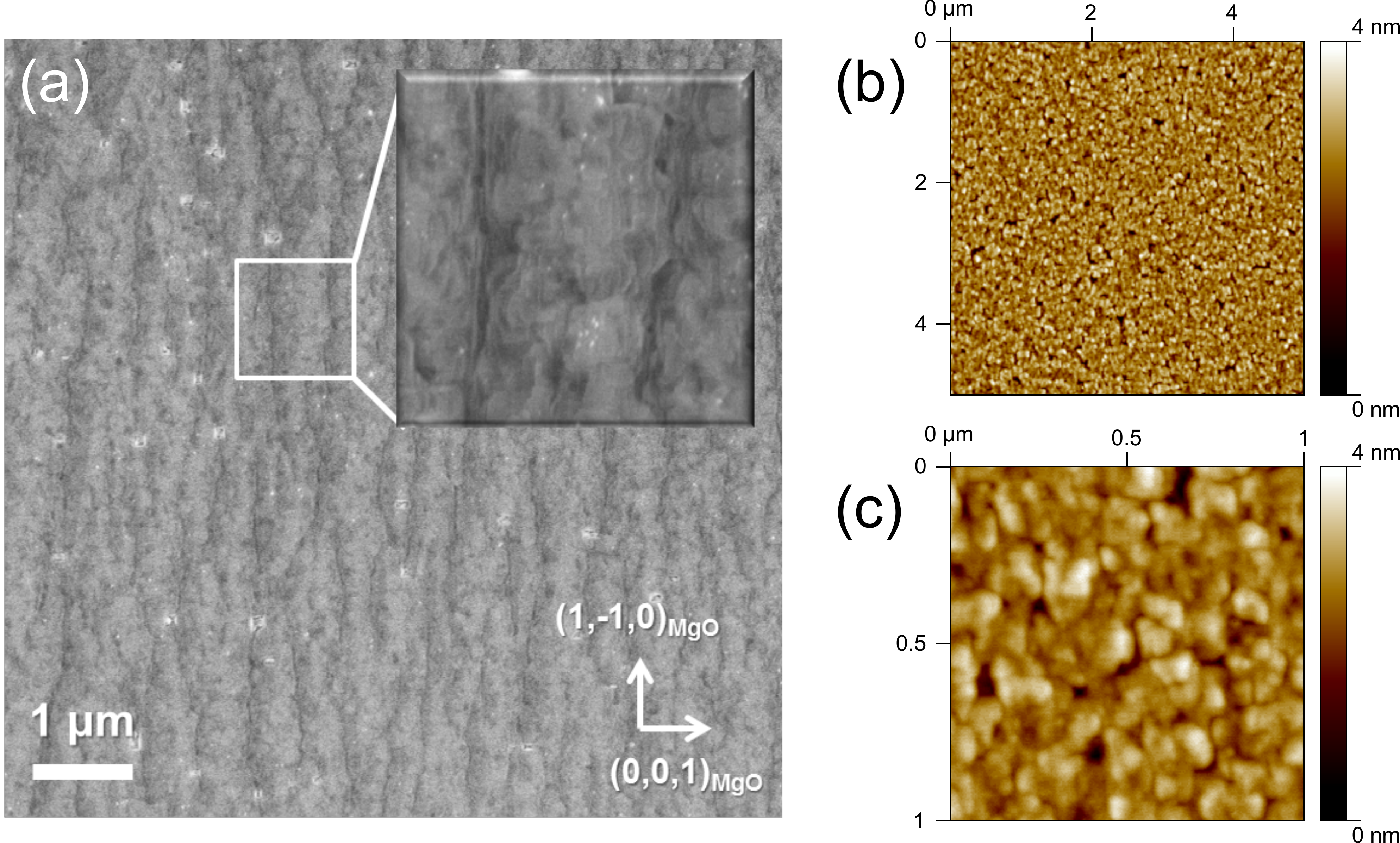}
\caption{(a) SEM picture of a 50 nm thick YBCO film. In the inset, elongated domains made of $c$-axis spirals are highlighed. (b)-(c) AFM pictures, at two different magnifications, of a 10 nm thick film. The roughness is $\approx 1.1$ nm, approximately equal to the thickness of a single atomic layer, even integrating on several micron large areas.}
\label{fig:SEMAFM}
\end{figure}

The surface morphology of the films has been studied by Atomic Force Microscopy (AFM) and Scanning Electron Microscopy (SEM) (see Fig. \ref{fig:SEMAFM}).

The films present smooth surfaces, characterized by the typical $c$-axis domains with 3D spirals (see Fig. \ref{fig:SEMAFM}(a)), and an average roughness which is in the range 0.8-1.2 nm ($\approx$ one atomic cell, see Fig. \ref{fig:SEMAFM}(b)-(c)).

\section{IVC characterization of 10 nm thick bare YBCO nanowires}

\begin{figure}[!ht]
\includegraphics[width=18cm]{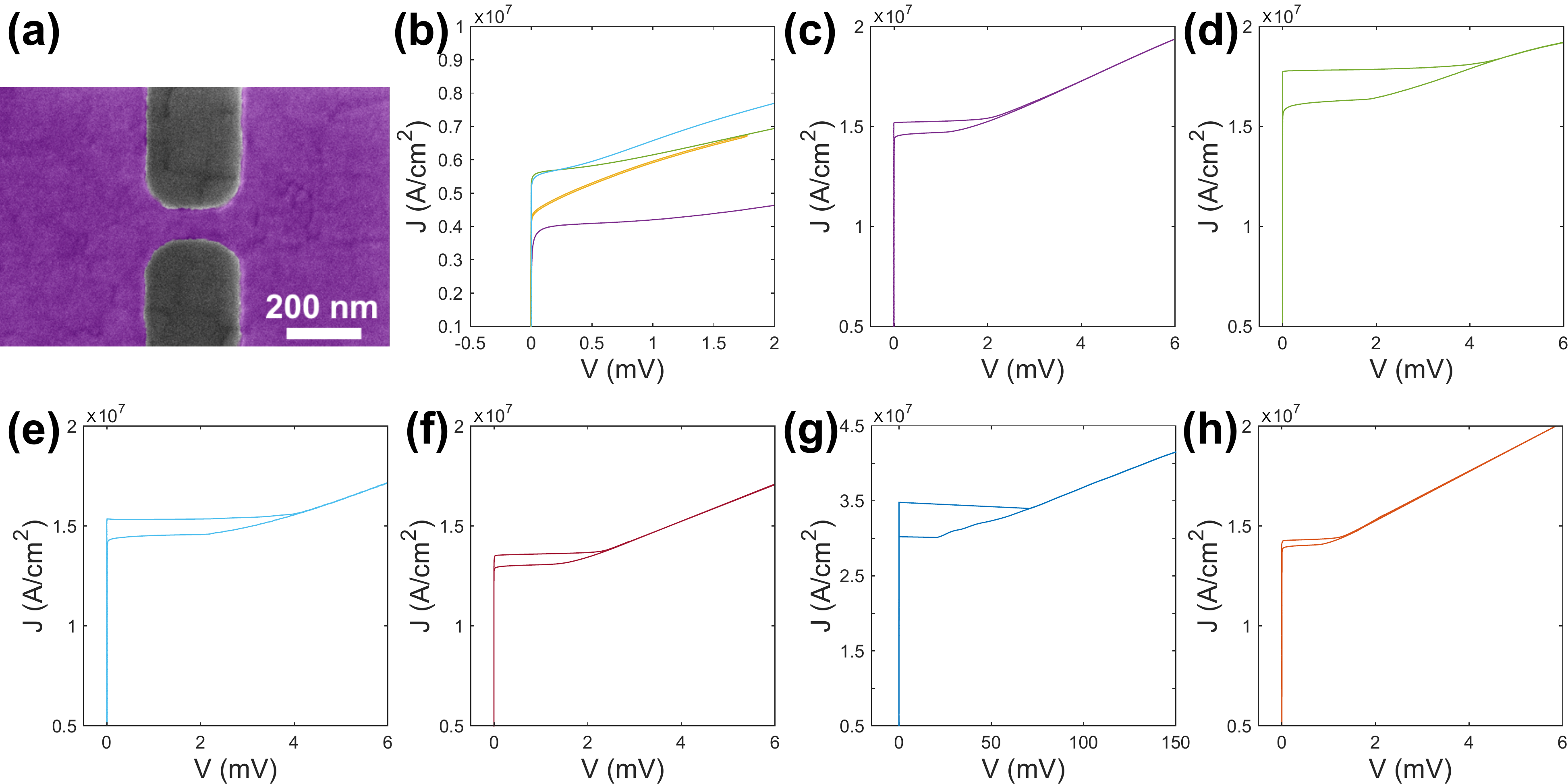}
\caption{(a) SEM image of a 100 nm long, 75 nm wide nanowire patterned on a 10 nm thick bare YBCO film. (b)-(h) $JV$ characteristics extrated by a distribution of wires. Wires having $J_C > 5 \cdot 10^6$ A/cm$^2$ present a voltage switch, of the order of several mV, above $I_C$, while nanowires with $J_C$ lower than this threshold are characterized by flux-flow like IVC.}
\label{fig:SEMJCV}
\end{figure}

We have patterned several bare YBCO films with a thickness $t = 10$ nm. Nanowires have been realized (see Fig. \ref{fig:SEMJCV}(a)), having widths $w$ in the range 65 - 120 nm, lengths $l$ in the range 100 - 500 nm, using the same nanopatterning procedure described in details in Ref. [\!\citenum{arpaia2013improved, nawaz2013approaching}].

The nanowires have been characterized by measuring the IV characteristics (IVC) at 4.2 K. Because of the substitution of the Au capping layer with a carbon protective layer during the nanopatterning we obtain lower $\bar{J}_C$ values and a larger spread of $J_C$ values with respect to the Au capped nanowires, used as reference systems. 

We measured in total 38 nanowires, 32 of which present a voltage switch from the superconducting to the normal state. In the following we show a set of IV curves measured on nanowires patterned on two chips, presenting $J_C$ both lower (IV without switch, see Fig. \ref{fig:SEMJCV}(b)) and higher (IV with a switch, see Fig. \ref{fig:SEMJCV}(c)-(h))  than $5\cdot10^6$ A/cm$^2$. 

\end{document}